\documentclass[final,1p,times,11pt,a4paper]{article}

\usepackage{hyperref}  
\hypersetup{pdfpagemode=None,pdfstartview=FitH}
\usepackage{fullpage}
\usepackage{amsmath}
\usepackage{amsbsy}
\usepackage{amssymb}
\usepackage{amsthm}
\usepackage{amscd}
\usepackage{amsfonts}
\usepackage{supertabular}
\usepackage{graphicx}
\usepackage{verbatim}
\usepackage{subfigure}
\usepackage{epsfig}
\usepackage{xspace}
\usepackage{euscript}
\usepackage{alltt}
\usepackage{boxedminipage}
\usepackage{float}
\usepackage{dsfont}
\usepackage{mathrsfs}
\usepackage[numbers]{natbib}
\usepackage{color}
\usepackage{txfonts}
\usepackage{upgreek}
\usepackage{bm}
\usepackage{authblk}
\usepackage[font=normal,labelfont=bf]{caption}
\usepackage[a4paper, total={6.5in, 9.5in}]{geometry}
\usepackage{lineno, blindtext}
\usepackage[normalem]{ulem}
\graphicspath{./Figures/pngfiles/}

\newcommand{\invisible}[1]{}
\graphicspath {
    {./}
    {./Figures/pngfiles/}
}

\begin{document}

\title{Peeling dynamics of fluid membranes bridged by molecular bonds: moving or breaking}

\author[a]{Dimitri Kaurin}
\author[a]{Pradeep K.~Bal}
\author[a,b,c]{Marino Arroyo\thanks{Corresponding author: marino.arroyo@upc.edu}} 

\affil[a]{Universitat Polit\`ecnica de Catalunya-BarcelonaTech, 08034 Barcelona, Spain}
\affil[b]{Institute for Bioengineering of Catalonia (IBEC), The Barcelona Institute of Science and Technology, 08028 Barcelona, Spain}
\affil[c]{CIMNE}

\date{}
\setcounter{Maxaffil}{0}
\renewcommand\Affilfont{\itshape\small}

\maketitle


\begin{abstract}
\normalsize
\noindent Biological adhesion is a critical mechanical function of complex organisms operating at multiple scales. At the cellular scale, cell-cell adhesion is remarkably tunable to enable both cohesion and malleability during development, homeostasis and disease. Such adaptable adhesion is physically supported by transient bonds between laterally mobile molecules embedded in fluid membranes. Thus, unlike specific adhesion at solid-solid or solid-fluid interfaces, peeling at fluid-fluid interfaces can proceed by breaking bonds, by moving bonds, or by a combination of both. How the additional degree of freedom provided by bond mobility changes the mechanics of peeling is not understood. To address this, we develop a theoretical model coupling self-consistently diffusion, reactions and mechanics. Lateral mobility and reaction rates determine distinct peeling regimes. In a diffusion-dominated Stefan-like regime, bond motion establishes self-stabilizing dynamics that increase the effective adhesion fracture energy. A reaction-dominated regime exhibits traveling peeling solutions where small-scale diffusion and marginal unbinding control peeling speed. In a mixed reaction-diffusion regime, strengthening by bond motion competes with weakening by bond breaking in a force-dependent manner, defining the strength of the adhesion patch. In turn, patch strength depends on molecular properties such as bond stiffness, force sensitivity, or crowding. We thus establish the physical rules enabling tunable cohesion in cellular tissues and in engineered biomimetic systems.  
\end{abstract}

\clearpage

\section*{Introduction}

{C}ell-cell adhesion is an essential mechanical function required to maintain  tissue integrity under mechanical stress  \cite{harris2012characterizing,Casares2015a,Latorre:2018aa}, disrupted during cancer \cite{Friedl:2011aa}, and finely tuned during development \cite{lecuit2007cell,guillot2013mechanics,Dumortier} or wound healing \cite{Tetley:2019aa}. Cell-cell adhesion needs to manage a contradiction between stability and malleability. Such versatile and adaptable interfaces avoid unspecific adhesion \cite{Bell1984a}, and instead rely on the collective effect of weak transmembrane bonds, notably of the cadherin family \cite{Yap2015}. Cell-cell adhesion is a multi-scale and highly regulated function that involves bond clustering, coupling to the cytoskeleton through mechanosensitive adapter proteins, turnover through endocytosis \cite{Yap2017,maitre2013three}, and Ca$^{2+}$-mediated control of the molecular properties of the binders and bonds, including diffusivity \cite{Yap2015,Cai2016}, stiffness \cite{Cai2016}, or force-sensitivity \cite{Rakshit2012}. The distinguishing physical feature of cell-cell adhesion as compared to cell-matrix adhesion, and in general compared to conventional specific adhesion at solid-solid or solid-fluid interfaces, is that both free binders and bonds are embedded in fluid membranes and hence are laterally mobile. As a result, the dynamics of adhesion between cells, and more generally between fluid membranes bridged by transient bonds, depend on binding/unbinding reactions between partner molecules and on the lateral motion bonds and free binders. Despite this fact has been long acknowledged \cite{Bell1984a,Dembo1988,Brochard-Wyart2002a,DeGennes2003}, its consequences on the dynamics of peeling are not understood and a mapping of the different dynamical scenarios of decohesion is lacking. 

To isolate the physical aspects of cell-cell adhesion, previous studies have focused on minimal artificial  model based on lipid membranes decorated with adhesion molecules \cite{Albersdorfer1997,Zhu2000,Nam2007,Smith2008,Sackmann2014,Schmid2016,fenz2017membrane}. While the theoretical understanding of equilibrium in such systems is established \cite{Bell1984a,Maitre2012a}, adhesion dynamics under force have been barely studied theoretically even though the mechanical environment of cell-cell adhesions is fundamentally dynamical. Here, we focus on the dynamics of unbinding of two vesicles held together by an adhesive patch made of mobile adhesion molecules forming transient \emph{trans} bonds. Our model is also pertinent to the forced unbinding of adjacent cells, since adhesion molecules attached to the cortex are still mobile due to turnover of cortical components. 

During peeling, an adhesion patch shrinks possibly until complete separation. In a tear-out limiting scenario, shrinkage of the patch may proceed by sequential bond breaking \cite{Evans1985,Evans1985a,Dembo1988,Berk1991,Pierrat2004,Lin2007a,Brochard-Wyart2002a}. When a vesicle with mobile binders adheres to a substrate with fixed receptors, spreading critically depends on diffusion of free binders on the vesicle \cite{Freund2004a,shenoy2005growth,Gao9469}, but bonds being immobile, peeling necessarily proceeds by tear-out. Peeling by bond breaking has been extensively studied theoretically \cite{Dembo1988,e2007,Gao2011,PhysRevLett.123.228102}. In a competing limiting scenario, the patch may shrink by lateral motion of bonds leading to an increasingly crowded patch \cite{Brochard-Wyart2002a,DeGennes2003}, a situation observed during cell-cell separation in vitro and in developing embryos \cite{10.1083/jcb.98.4.1201,Tozeren1989,maitre2012adhesion,Dumortier}. In general, the dynamics of decohesion may depend on a combination of bond breaking and bond motion (or reaction and diffusion), but this interplay has not been systematically examined \cite{Evans1985a,Dembo1988,Zhu1991,Brochard-Wyart2002a} despite the fact that bond mobility has been shown to strongly influence adhesions in hybrid cell-supported bilayer studies \cite{BISWAS2020163} and in purely artificial systems \cite{Smith2008}.

To understand the physical principles governing peeling of adhesive interface bridged by mobile bonds, we develop a self-consistent continuum dynamical model capturing the reaction kinetics of bond formation and dissociation, the lateral diffusion of adhesion molecules, and the mechanics of the adhesion patch and of the adhering vesicles. We then identify and characterize distinct regimes pertinent to (1) long-lived mobile bonds, (2) short-lived bonds with reduced mobility such as cadherin molecules linked to the cell cytoskeleton, and (3) short-lived mobile bonds such as cadherins in a lipid bilayer. Each of these regimes exhibit fundamentally different dynamics (self-similar, traveling or mulitphasic) with multi-scale features in space and time. We further examine the relation between molecular properties of bonds  and the effective behavior of the adhesive patch.

\section*{Methods}

\subsection*{Theoretical and computational model}

\begin{figure}
\centering
  \includegraphics[width=.65\linewidth]{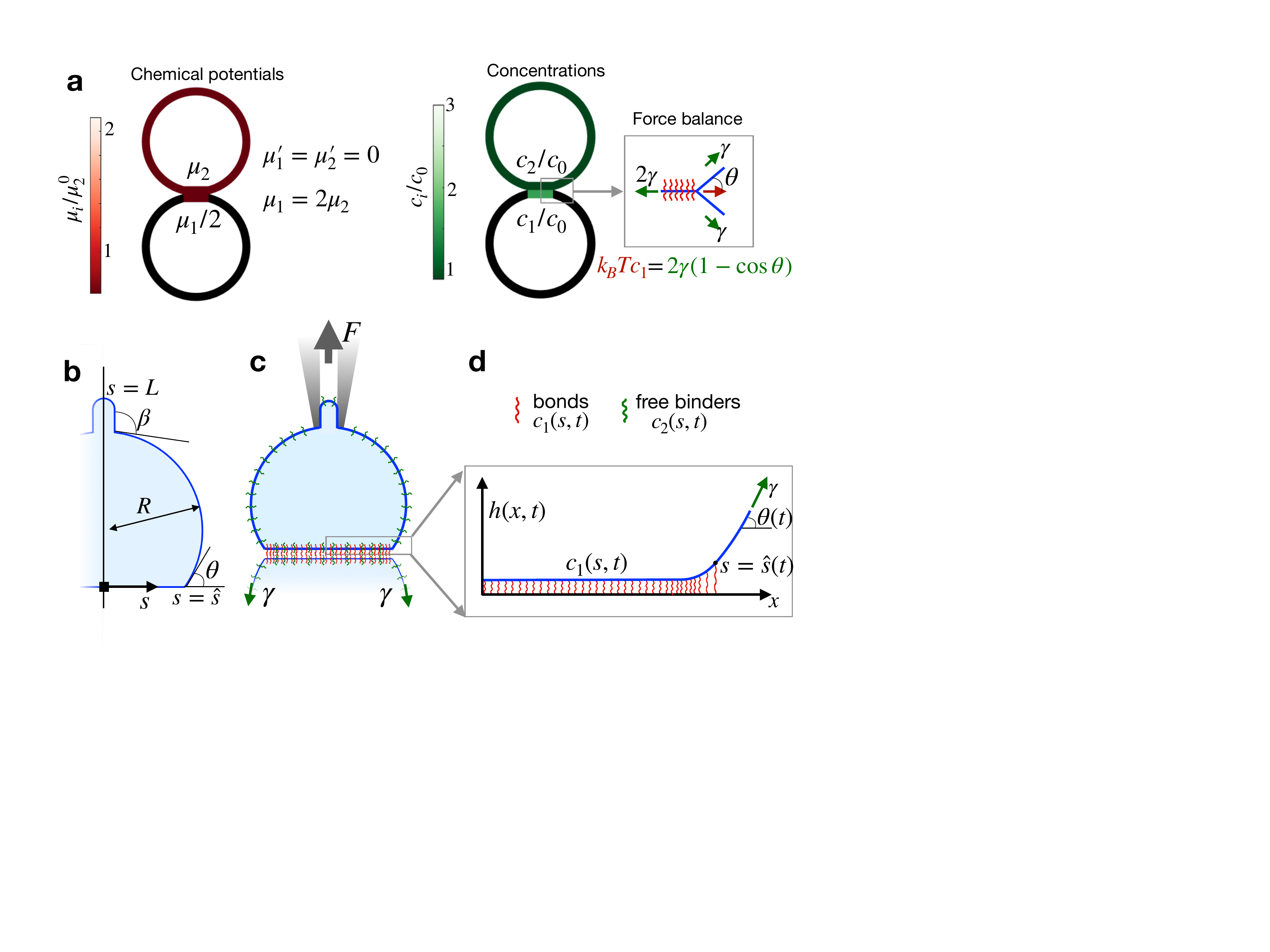}
  \caption{(a) Equilibrium of two identical vesicles adhering through mobile non-compliant binders. Chemical equilibrium requires uniformity and equality of the chemical potentials (left). Mechanical equilibrium requires satisfaction of Young-Laplace and Young-Dupr\'e relations (right). (b) Capillary model describing membrane mechanics in terms of the half-size of the adhesion, $\hat{s}$, the  angles $\theta$ and $\beta$, and the radius $R$.  (c) Schematic of the system, where a loading device  controls surface tension $\gamma$ and  force $F$. (d) Micro-mechanics model of the adhesion patch resolving the separation profile relative to an equilibrium separation, $h$, accounting for bending stiffness and bond compliance. }
  \label{fig1}
\end{figure}

The state of the system is defined  by the shape of the adhering vesicles, the number concentration of bonds on the adhesion patch, $c_1$, and that of free binders on the entire vesicle, $c_2$. We assume that the adhering vesicles are identical and are made  of a fluid membrane where bonds and free binders are mobile. Focusing first on a dilute limit and non-compliant bonds, the chemical potentials of bonds and free binders take the form $\mu_i = \mu^{0}_i+ k_BT \log {c_i}/{c_0}, ~i=1,2$, where $\mu^{0}_i$ is the standard chemical potential, $k_BT$ is the thermal energy scale, and $c_0$ is a reference concentration. In equilibrium, shape and concentrations obey chemical and mechanical equilibrium conditions \cite{Maitre2012a}. Chemical equilibrium requires that $\mu_1$ (and hence $c_1$) is uniform on the adhesion patch, that $\mu_2$ (and hence $c_2$) is uniform on the entire vesicle including the adhesion patch, and that $\mu_1 = 2\mu_2$ over the patch since two free binders react to form a bond. This last condition implies that
\begin{align}\label{eqK}
K= c_0 c_1/c_2^2 = \exp \left[({2\mu_2^0 - \mu_1^0})/{k_BT}\right].
\end{align}
with $K$ the equilibrium constant. At high membrane tension $\gamma$ expected for a mature adhesion patch rich in bonds, the length-scale $\ell_1 = \sqrt{\kappa/\gamma}$, where $\kappa$ is the bending stiffness, is much smaller than the vesicle size. In this capillary limit, mechanical equilibrium at the edge of the adhesion patch is formally a Young-Dupr\'e equation $k_BT c_1 = 2\gamma (1-\cos\theta)$, where the left-hand side is the osmotic tension of the bonds in the adhesion patch and $\theta$ is the contact angle between the free part of the vesicle and the symmetry plane, Fig.~\ref{fig1}a. Besides justifying the capillary limit, high tension suppresses thermal fluctuations of the membrane \cite{Parolini:2015aa}, which are not considered here but play a prominent role during low-tension adhesion spreading \cite{Smith2008,fenz2017membrane}. Mechanical equilibrium in the non-adhered part of the vesicle is expressed by Laplace's law. These conditions determine the equilibrium shape of the vesicles, the size of the adhesion and the concentration of free binders and bonds. 

Describing the out-of-equilibrium dynamics under force requires accounting for diffusion and chemical kinetics, which in turn depend on mechanics in different ways. Diffusion of bonds is biased by their tendency to leave regions where they are highly stretched. Chemical kinetics are influenced by mechanics since unbinding rates are force sensitive and  rebinding rates depend on the distance between potential partners \cite{e2007,PhysRevLett.123.228102}. We thus need to resolve the force born by bonds and the separation between membranes required to compute these rates. Exploiting scale separation, we combine a vesicle-scale capillary model with a model for the micro-mechanics of the adhesion patch accounting for the bending rigidity of the membrane and the compliance of the molecular bonds, Fig.~\ref{fig1}c,d. In this model, the length-scale over which the tension of the free-standing membrane is transmitted to the adhesion patch can be estimated as $\ell_2 = \sqrt[4]{\kappa/(k c_0)}$, where $k$ is the bond stiffness and $c_0$ is a typical bond concentration. For reasonable  parameters, $\ell_1$ and $\ell_2$ are in the order of 20 nm, much smaller than the typical size of an adhesion patch.

To focus on the mechano-chemistry of forced decohesion and simplify all other aspects of the model, we restrict ourselves to a 2D geometry where the membrane becomes a line whose arc-length coordinate is denoted by $s$. We summarize here the theoretical model and provide a detailed derivation in Supplementary Note 1. The vesicle is connected to a loading device simulating a micropipette, which controls membrane tension $\gamma$ by drawing or supplying length and applies a vertical force $F$, Fig.~\ref{fig1}c \cite{10.1083/jcb.98.4.1201,Tozeren1989,maitre2012adhesion}. Mechanical relaxation is much faster than chemical relaxation, and thus we treat mechanics quasi-statically. Given the prescribed $\gamma$ and $F$ and the current size of the adhesion patch, $\hat{s}(t)$, the capillary model provides the shape of the vesicle,   Fig.~\ref{fig1}b, in particular contact angle $\theta$. With this information along with the current concentration of bonds $c_1(s,t)$, the micro-mechanical model provides the membrane separation profile $h(s,t)$, Fig.~\ref{fig1}d, which allows us to compute the out-of-plane force per molecule $k h(s,t)$ within the patch region. In turn, this information allows us to evolve the bond and free-binder concentrations $c_1(s,t)$ and $c_2(s,t)$ and the position of the interface $\hat{s}(t)$ as discussed next.

The reaction-diffusion dynamics for $c_1$ and $c_2$ is given by
\begin{alignat}{4}
\dot{c}_1 & = D_1 \left[ c_1' + c_1 \left({h^2}/{x_\gamma^2}\right)'  \right]'  &+ k_{\rm on} c_2^2 - k_{\rm off} c_1 & \;\;\mbox{in } (0,\hat{s}(t)) \label{GEqn1} \\
\dot{c}_2 & = D_2 c_2''  &- k_{\rm on} c_2^2 + k_{\rm off} c_1 & \;\; \mbox{in } (0,\hat{s}(t)) \label{GEqn2}\\
\dot{c}_2 & = D_2 c_2'' & & \;\;\mbox{in } (\hat{s}(t),L_0) \label{GEqn3}
\end{alignat}
where dots and primes denote time and space derivatives, $D_{1,2}$ are diffusion constants of bonds/free-binders, $k_{\rm on}$ is the binding rate, $k_{\rm off}$  the unbinding rate, $L_0$ is the total membrane length, and $x_\gamma = \sqrt{k_BT/k}$ the scale of thermal fluctuations of binders. These partial differential equations are defined on a time-dependent domain. The transport term in Eq.~(\ref{GEqn1}) includes a diffusive term and a bias, according to which bonds try to reduce the mechanical contribution of their chemical potential \cite{Zhu1991}
\begin{align} \label{chem2}
\mu_1(c_1,h) = \mu^{0}_1+ k_BT \log \frac{c_1}{c_0} + k_BT \left({h}/{x_\gamma}\right)^{2},
\end{align}
by moving away from regions where they are highly stretched.  In general, the binding and unbinding rates depend on the separation between the adjacent membranes. We recover the classical expression for the binding rate \cite{e2007,PhysRevLett.123.228102}
\begin{align}\label{kon}
k_{\rm on}(h) = \bar{k}_{\rm on} \exp\left[ -\left({h}/{x_\gamma} \right)^2 \right],
\end{align}
expressing that a bond is more likely to form if  potential partners are close or if thermal fluctuations are large. Although cadherin bonds are thought to shift between ideal, slip or catch bonds depending on environmental conditions and conformation \cite{Rakshit2012,Manibog2014}, here we only consider the slip-bond behavior as described by Bell's model \cite{bell1978models}
\begin{align}\label{koff}
k_{\rm off}(h) = \bar{k}_{\rm off} \exp\left({k h}/{f_\beta} \right)
= \bar{k}_{\rm off} \exp\left({h}/{x_\beta} \right),
\end{align}
where $f_\beta$ is the force sensitivity and where we introduce a separation sensitivity $x_\beta = f_\beta/k$ for convenience.

The governing equations (\ref{GEqn1}-\ref{GEqn3}) need to be supplemented by initial, boundary and interface conditions at $s = \hat{s}(t)$. Since free binders can cross the interface, their concentration and flux are continuous.  
In contrast, the interface is by definition a barrier for bonds. Consequently the diffusive flux of bonds at the interface must be compensated by bond transport due to interface motion, 
\begin{align} \label{flux_int}
-D_1 \left\{ c_1' + c_1 \left[\left({h}/{x_\gamma}\right)^2\right]'  \right\}_{s = \hat{s}^-} = c_1(\hat{s}^-,t) \hat{v},
\end{align}
where $\hat{v} = d\hat{s}/dt$ is the velocity of the interface. Finally, force balance at the interface requires that 
\begin{align} \label{mech_int}
k_BT c_1 (\hat{s}^-,t) = 2\gamma (1-\cos\theta).
\end{align}
Comparison of this equation with Rivlin's classical theory of peeling \cite{Rivlin1997} shows that osmotic pressure of bonds at the interface, $k_BT c_1(\hat{s}^-,t)$,  plays the role of the adhesion fracture energy. However, rather than a material property of the interface as in classical peeling and in the case of tear-out of a vesicle against a substrate with immobile receptors  \cite{DeGennes2003}, here this quantity is a dynamical variable.
Equations (\ref{GEqn1}-\ref{GEqn2},\ref{flux_int},\ref{mech_int}) supplemented by the initial  and boundary conditions at $s=0$ and $s=L_0$ allow us to solve for ${c}_1(s,t)$, ${c}_2(s,t)$ and $\hat{s}(t)$. 
As these variables evolve, we need to update the mechanical variables $\theta$ and $h(s,t)$, which in turn affect the reaction-diffusion-interface dynamics. The self-consistent finite element numerical solution of the model is described in detail in Supplementary Note 1.

\subsection*{System preparation and parameters}
Before driving the system out-of-equilibrium, we prepared the system at an equilibrium state for non-compliant and ideal bonds. We set $K= 2$, $F=0$, the vesicle radius to $R_0= 10$~$\upmu$m, $\hat{s}_0= 2.5$~$\upmu$m, $c_0 = 2.5\cdot10^3$~molecules /$\upmu$m$^2\times \ell_{\rm lat}$, and the length of half vesicle to $L_0=35$~$\upmu$m. Thus, the total number of molecules is  $N_{\rm tot} = c_0 \times L_0$. Here, $\ell_{\rm lat}$ is an arbitrary depth of our 1D membrane to make it ribbon, allowing us to map 2D to 1D number densities. Without loss of generality, $\mu_1^0=0$ and from Eq.~(\ref{eqK})  conclude that $\mu_2^0 = (k_BT\log K)/2$. In all figures, we non-dimensionalize chemical potentials by $\mu_2^0$. With these data, conservation of the total number of molecules, $N_{\rm tot} = c_1 \hat{s}_0 + c_2 L_0$ and the law of mass action $c_0 c_1/c_2^2 = K$ provide two equations to solve for the equilibrium concentrations, obtaining  $c_1 = 1.58 c_0$ and $c_2= 0.89 c_0$. Since $R_0$ and $\hat{s}_0$ determine the contact angle as $\sin\theta_0 = \hat{s}_0/R_0$, force balance at the interface provides an equation for the membrane tension, for which taking $k_BT = 4.11\cdot 10^{-21}$ J we find $\gamma=2.55\cdot10^{-4}$~N/m $\times \ell_{\rm lat}$. This equilibrium state is illustrated in Fig.~\ref{fig1}a.

Starting from this state and  driving it out-of-equilibrium by suddenly increasing the applied force $F$, we tracked the mechano-chemical dynamics of forced decohesion, which tend to uniformize $\mu_{1}$ and $\mu_{2}$  over the patch and vesicle, and tend to equilibrate them as $\mu_1 = 2\mu_2$ over the patch. The time-scale of bond diffusion in the adhesion patch is $\tau_{\rm diff,1}={\hat{s}_0^2}/{D_1}$, whereas that of free-binders on the entire vesicle is $\tau_{\rm diff,2}={L_0^2}/{D_2}$. Since a bond connects two binders, in the simplest situation its mobility is half of that of a free-binder, hence $D_2 = 2 D_1$. With our choice of parameters, $\tau_{\rm diff,2} \approx 100 \,\tau_{\rm diff,1}$. Regarding reactions, the natural time-scale is $\tau_{\rm reac}= 1/{\bar{k}_{\rm off}}$. Once ${\bar{k}_{\rm off}}$ is fixed, we determined the binding rate from $K/c_0 =\bar{k}_{\rm on}/\bar{k}_{\rm off}$, which results from Eqs.~(\ref{eqK},\ref{GEqn1}). The ratios between the reaction time-scale and the diffusive time-scales introduce Damkohler numbers weighing the relative importance of reactions and diffusion. For compliant bonds/binders, we consider a reference stiffness of $k=2.5\cdot 10^{-4}$~N/m.

\section*{Results}

\subsection*{Diffusion-dominated regime}

\begin{figure*}
\centering
  \includegraphics[width=.99\linewidth]{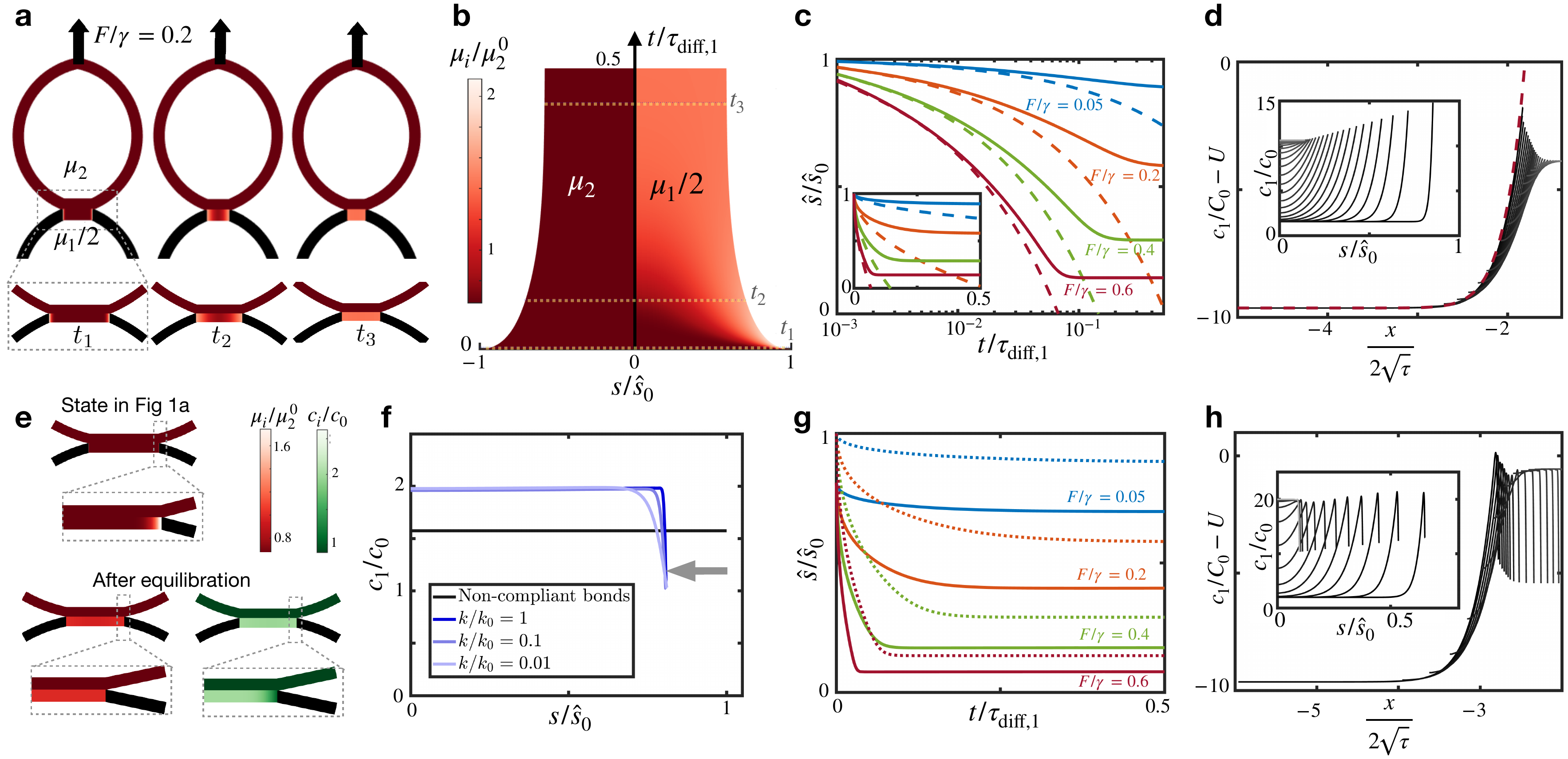}
  \caption{Diffusion-dominated regime for non-compliant (a-d) and compliant (e-h) bonds. (a) Snapshots of shape of the adherent vesicles and the chemical potentials after application of $F/\gamma=0.2$ with non-compliant bonds. (b) Kymograph of interface position and  chemical potentials of free binders (left) and bonds (right). (c) Position of the interface as a function of time for different applied forces. Dashed lines are the short-time analytical predictions in Eq.~(\ref{exact2}). Semi-logarithmic scale highlights the short-time behavior, see inset for linear scale. (d) Bond concentration as a function of position at various instants (inset) and in terms of the similarity variable; the red dashed line is  the analytical solution in Eq.~(\ref{exact1}). (e) The state in Fig.~\ref{fig1}a is not in equilibrium for compliant bonds since stretching near the interface of the patch  increases $\mu_1$. Upon equilibration $\mu_1$ becomes uniform but cannot equilibrate with $\mu_2$ in the absence of reactions, and $c_1$ has a boundary layer depleted of bonds. (f) Dependence of the equilibrium bond distribution and patch size on bond stiffness normalized by $k_0=2.5\cdot 10^{-4}$~N/m. (g) Dynamics of the interface upon force application with compliant bonds; dotted lines indicate the interface dynamics for non-compliant bonds. In (b,c,f,g), $\hat{s}$ is nondimensionalized by the equilibrium size of the patch at $F=0$ and non-compliant bonds. (h) Bond concentration for compliant bonds as a function of $s$ (inset) and as a function of the similarity variable.}
  \label{fig2}
\end{figure*}

We first focused on the situation in which reactions are extremely slow compared to diffusion by setting $\bar{k}_{\rm on}$ and $\bar{k}_{\rm off}$ to zero and thus $\tau_{\rm reac}= +\infty$. In this limit, the dynamics of free binders become uncoupled to the rest of the system. Thus, we are only left with one time-scale associated to diffusion of bonds. For typical diffusion coefficient of free-binders on lipid membranes, $D_2 = 2D_1= 0.5$~$\upmu$m$^2$/s, this time-scale is  $\tau_{\rm diff,1}\approx 25$ s.  We first considered non-compliant bonds by setting $h(s,t)=0$, see Supplementary Note 1. For this model, the state in  Fig.~\ref{fig1}a is an exact equilibrium state at $F=0$. We then applied suddenly a separation force $F$. We reasoned that in response to force application, the contact angle should rapidly increase $\theta>\theta_0$, requiring a concomitant increase in bond concentration at the interface according to Eq.~(\ref{mech_int}). In turn, this should create a sharp positive gradient in bond concentration and chemical potential, and hence a motion of the interface leading to patch shrinkage,  Eq.~(\ref{flux_int}).  As the patch becomes smaller, bond chemical potentials should equilibrate faster since they diffuse over a smaller distance and, since the number of bonds is constant, the patch should become increasingly concentrated with bonds. Our simulations confirmed this physical picture, Fig.~\ref{fig2}a,b, Suppl.~Fig.~1a,c and Suppl.~Movie 1, and the system reaches a new equilibrium state where $\mu_1$ is high and uniform and the higher bond osmotic tension balances the larger out-of-equilibrium membrane force at the interface, Eq.~(\ref{mech_int}). Hence, the system is able self-adjust the effective adhesion fracture energy, $k_BT c_1 (\hat{s}^-,t)$, to balance the higher peeling driving force. Although the final chemical potential of bonds is much larger than that of free-binders, equilibration between species is not possible in the absence of reactions, Fig.~\ref{fig2}a,b.

As $F$ increases, $\theta$ also increases and a larger osmotic tension, and hence  a larger bond concentration, is required to balance the mechanical force at the interface. Since the number of bonds is constant, this in turn requires a smaller equilibrium patch, Suppl.~Fig.~1b, Fig.~\ref{fig2}c. Interestingly, a similar process of shrinkage and concentration of adhesive patches has been observed in cell doublets under force in vitro, and during cell-cell hydraulic fracture in developing embryos \cite{Maitre2012a,Dumortier}.

Increasing the force also leads to faster dynamics, Fig.~\ref{fig2}c, Suppl.~Movie 1. To further understand peeling dynamics, we sought  an analytical solution. Starting from the equilibrium state in Fig.~\ref{fig1}a with contact angle $\theta_0$, patch size $\hat{s}_0$ and uniform bond concentration $c_1(s,0) = C_0 = 2\gamma(1-\cos\theta_0) /(k_BT)$, we suddenly increased force and hence $\theta>\theta_0$. In the actual system, as $\hat{s}(t)$ decreases so does $\theta(t)$ due to vesicle capillarity, Fig.~\ref{fig2}d(inset). To develop the analytical solution,  we simplified the problem by assuming that $\theta$ remains constant during peeling. Introducing non-dimensional space $x = s/\hat{s}_0-1$, time $\tau=t/\tau_{\rm diff,1}$, driving parameter $U=(1-\cos\theta)/(1-\cos\theta_0)$,  position of the interface $X(\tau) = \hat{s}(\tau_{\rm diff,1}\tau)/\hat{s}_0-1$ and  concentration of bonds  $u(x,\tau) = c_1(\hat{s}_0 (x+1),\tau_{\rm diff,1}\tau)/C_0 - U$, the governing Eqs.~(\ref{GEqn1},\ref{flux_int},\ref{mech_int}) take the form of a classical Stefan problem, see Supplementary Note 2, which describes a myriad of phase transformation problems \cite{Gupta_stefan}. This problem admits an analytical solution valid at short times or for large domains in terms of the similarity variable $x/\sqrt{\tau}$,
\begin{align}
u(x,\tau)  &= \sqrt{\pi} \lambda  e^{\lambda^2} U\left[{\rm erf}(\lambda) -  {\rm erf}\left(\frac{x}{2\sqrt{\tau}}\right)\right], \label{exact1}\\ X(\tau) &=  2\lambda\sqrt{\tau},\label{exact2}
\end{align}
where ${\rm erf}$ is the error function and $\lambda$ is a constant depending on the driving parameter $U$ and implicitly determined by $\sqrt{\pi} \lambda  e^{\lambda^2} U\left[{\rm erf}(\lambda) +1\right] = 1-U$. To linear order in $U-1$, $\lambda(U)\approx -(U-1)/\sqrt{\pi} \propto (\cos\theta-\cos\theta_0)$, showing how the interface motion depends on the horizontal force imbalance. Interestingly, the diffusion-controlled spreading of a membrane with mobile molecules binding to fixed receptors exhibits analogous self-similar dynamics \cite{Freund2004a,shenoy2005growth}.

We then compared these analytical predictions, valid at short-times/large-domains and assuming that $\theta$ remains constant, with our numerical calculations, which did not make any of these assumptions. At short times, Eq.~(\ref{exact2}) predicts very well the motion of the interface, particularly at high forces where the interface moves significantly before the finite-size effects leading to self-stabilization of the interface start to play a role, Fig.~\ref{fig2}c. 
To further test the theory, we plotted the rescaled bond concentration at different time instants against the similarity variable, finding a remarkable collapse to Eq.~(\ref{exact1}) despite the reduction of $\theta$ with time and finite-size effects, Fig.~\ref{fig2}d. These results thus establish a mapping between the dynamics of forced adhesions between membranes mediated by long-lived bonds and the self-similar solution  of the classical Stefan problem.

Next, we examined numerically the more realistic situation of deformable bonds. In this case, the state in Fig.~\ref{fig1}a is not in equilibrium for $F=0$ since bonds are stretched in a boundary layer near the interface, which modifies their chemical potential, Eq.~(\ref{chem2}) and Fig.~\ref{fig2}e. Despite the small size of the perturbed region, the new equilibrium state is significantly smaller and more concentrated, Fig.~\ref{fig2}f and Suppl.~Fig 1c,d. The width of the boundary layer where bonds are stretched and depleted is commensurate to $\ell_2$ and thus decreases with increasing bond stiffness, but the new position of the interface is quite insensitive to $k$, Fig.~\ref{fig2}f. Interestingly, the results for non-compliant bonds are not the limit as $k\rightarrow +\infty$ of those for compliant bonds. Upon force application following equilibration, the dynamics proceed similarly to the case of non-compliant bonds, albeit at a faster rate, Fig.~\ref{fig2}g and Suppl.~Fig.~1c,d, Suppl.~Movie 2, which we ascribe to the advective bond transport due to gradients in bond stretching, Eqs.~(\ref{GEqn1},\ref{flux_int}). Although the similarity solution does not account for bond compliance, bond concentration at different times remarkably collapses when rescaled and plotted in the similarity variable, Fig.~\ref{fig2}h, showing that the Stefan problem captures the  physics of this diffusion-dominated regime at short times. 

We finally examined the force distribution in the adhesion patch. The out-of-plane traction  $k\, c_1(s,t) h(s,t)$ needs to balance $F$ and hence increases with it. However, increasing $F$ does not lead to further separation. Instead, the system adapts to higher $F$ (higher $\theta$) by increasing $c_1(s,t)$ near the interface, Eq.~(\ref{mech_int}), while keeping a rather constant profile of the force per molecule, $k\, h(s,t)$, with a force scale emerging from the competition of mixing entropy and bond stretching, Eq.~(\ref{chem2}), and given by $f_\gamma = \sqrt{k\cdot k_BT}\approx 1.0$ pN, Suppl.~Fig.~2.

\begin{figure*}[h]
\includegraphics[width=0.99\textwidth]{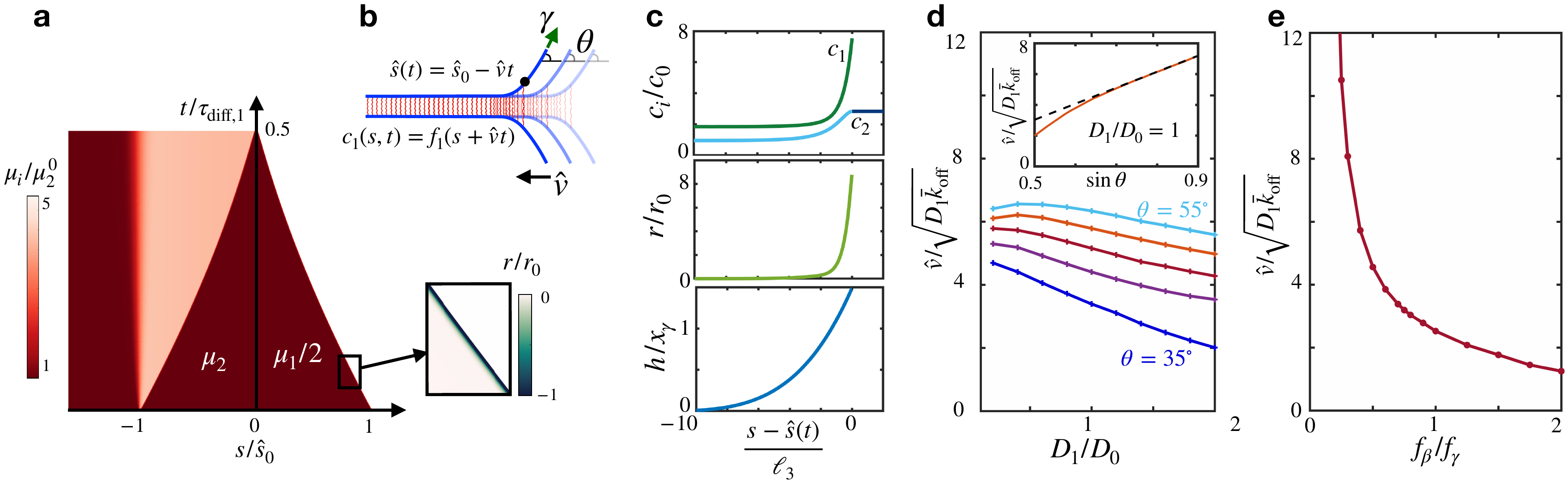}%
\centering
\caption{Reaction-dominated regime. (a) Kymograph of chemical potentials and net binding rate for $f_\beta/f_\gamma = 4$, $D_1/D_0=1$ and $F/\gamma = 0.5$  (inset), where $r_0 = \bar{k}_{\rm off} c_0$. (b) Illustration of the traveling solution where $\theta$ is kept constant, and (c) concentration, reaction rate and separation profiles for the traveling solution corresponding to  $f_\beta/f_\gamma = 4$, $D_1/D_0=1$ and $\theta = 32^\circ$. (d) Velocity of the interface normalized by the prediction $v_0 = \sqrt{D_1\bar{k}_{\rm off}}$ as a function of diffusivity for $f_\beta/f_\gamma = 4$ and several contact angles $\theta$. Normalized interface velocity as a function of $\sin\theta$ for $D_1/D_0 = 1$ (inset). (e) Normalized velocity of the interface as a function of force sensitivity for $D_1/D_0=1$ and $\theta = 32^\circ$.}
\label{fig4}
\end{figure*}

\subsection*{Reaction-dominated case}

We then studied a different extreme scenario characterized by fast reaction rates, $\bar{k}_{\rm off} = 10$ s$^{-1}$, representative of weak bonds such as cadherins, and very low diffusivity, which should result in a reaction-dominated regime similar to the tear-out of an adhesive vesicle from a solid substrate with immobile receptors \cite{Berk1991,Pierrat2004,Lin2007a}. We decreased $D_{1,2}$ by a factor of about  1000 ($ D_1 = D_2/2 = D_0 = 0.25\cdot 10^{-3}$ $\upmu$m$^2$/s), comparable to the reduction of diffusivity of adhesion molecules from artificial lipid bilayers to cell membranes \cite{Biswas10932}, and initially considered rather insensitive slip bonds with $f_\beta/f_\gamma = 4$. Upon force application, we observed that in contrast with the diffusion dominated case where $\hat{s}(t)-\hat{s}_0\propto \sqrt{t}$, now both the size of the patch and the number of bonds decrease nearly linearly, with the net unbinding reaction rate localized in the close vicinity of the interface, Fig.~\ref{fig4}a,  Suppl.~Movie 3. Complete decohesion is reached before significant diffusion of bond or free binders can take place at the patch or vesicle scales. 

In view of these results, we hypothesized the existence of traveling solutions of the form $c_i(s,t) = f_i(s+\hat{v}t)$, Fig.~\ref{fig4}b. To systematically examine this point, we again made the approximation of constant driving force (constant $\theta$). Propagating fronts in reaction-diffusion systems require non-generic nonlinearity,
as in the prototypical Fisher-Kolmogorov-Petrovskii-Piscunov (FKPP) equation \cite{KPP1937,fisher} or in the FitzHugh-Nagumo system \cite{Rinzel:1973aa}. For non-compliant ideal bonds, our model in the moving frame reduces to a two-species advection-reaction-diffusion equation whose only nonlinearity is the term $\bar{k}_{\rm on}c_2^2$, and our simulations did not develop traveling solutions. Instead, for compliant and force-sensitive bonds the model couples to mechanics, which predicts a separation profile $h$ localized near the adhesion edge that in turn biases bond motion away from the edge, locally increases off rates and decrease on rates.  In this case, our simulations readily developed traveling solutions with constant interface velocity, localized unbinding, a sharp transition of bonds to accommodate the concentration at $s=\hat{s}(t)$, Eq.~(\ref{mech_int}), and a sharp transition of free binders to a higher plateau in the wake of the interface due to broken bonds, Fig.~\ref{fig4}c, Suppl.~Fig.~3.

\begin{figure*}[h]
\includegraphics[width=0.99\textwidth]{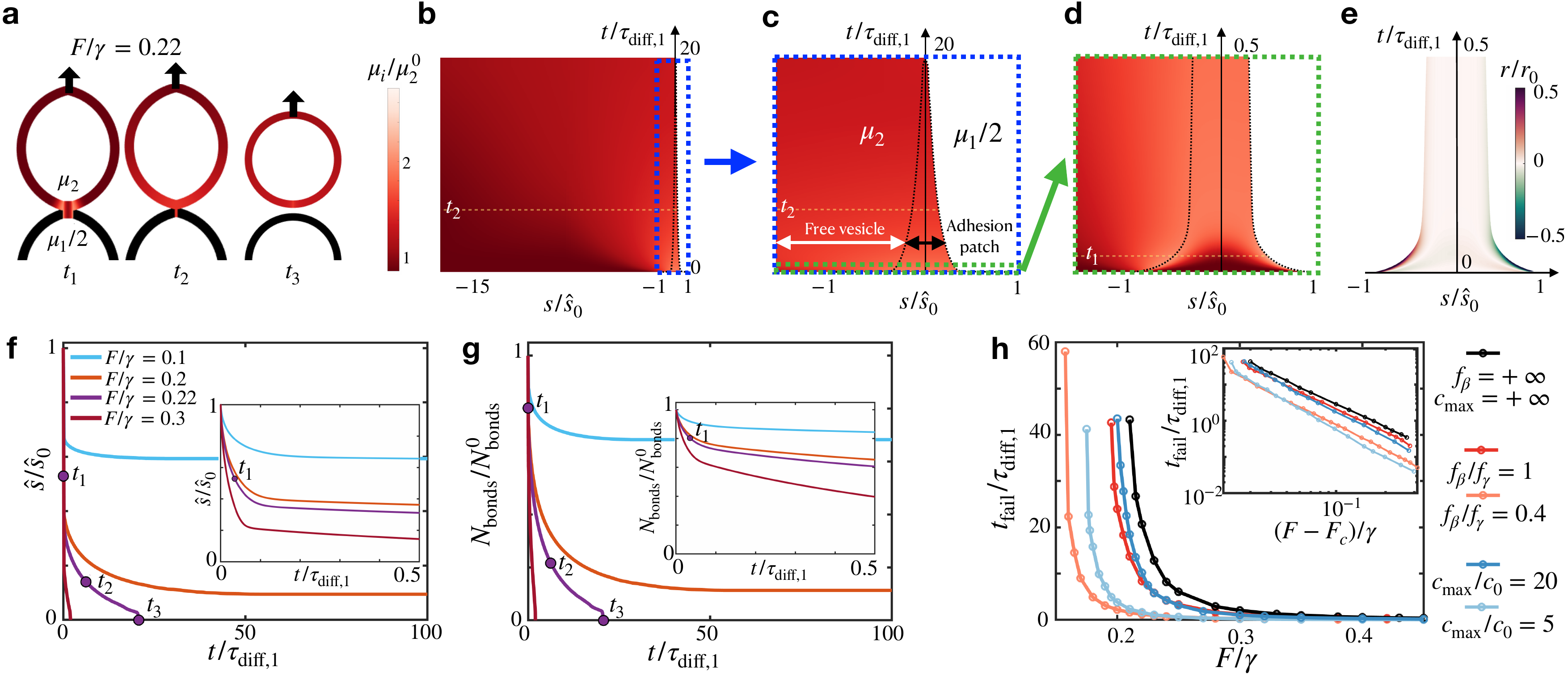}%
\centering
\caption{Mixed reaction-diffusion regime. (a) Snapshots of shape of the adherent vesicles and of the chemical potentials after application of $F/\gamma=0.22$ with compliant ideal bonds, leading to full decohesion. The three time-instants are labelled in panels b-g. (b-d) Kymographs at different space and time scales of the position of the interface (dashed black line) and the chemical potential of free binders over the entire vesicle (left) and bonds over the patch (right). (e) Kymograph of the net rate of free binder (left) and bond (right) increase due to reactions. (f) Position of the interface and (g) number of bonds in the patch as a function of time for different applied forces. (h) Failure time as a function of applied force for ideal and for slip bonds.}
\label{fig:mixed}
\end{figure*}

To understand the physics controlling the peeling speed $\hat{v}$ and the local profiles around $s = \hat{s}(t)$, we reasoned that even though the time of bond diffusion in the whole patch is very long, $\tau_{\rm diff,1} \gg \tau_{\rm reac}$,  there should be a small length-scale $\ell_3$ over which diffusion and reactions compete near the interface, $\ell_3^2/D_1 \approx \tau_{\rm reac}$, leading to $\ell_3 = \sqrt{D_1/\bar{k}_{\rm off}}\approx 5$ nm. For our parameter set, $\ell_3$ is smaller than $\ell_2$ over which bonds are loaded. Dimensional analysis and the analogy with the FKPP model suggest that this small-scale diffusion may control the overall decohesion process by setting the front speed $v_0 = \sqrt{D_1 \bar{k}_{\rm off}}\approx 50$ nm/s. This situation is also analogous to fracture mechanics of solids, where the physics within a small \emph{process zone} in the vicinity of the crack tip determine the effective fracture properties \cite{TVERGAARD19921377}. Our simulations confirmed that $v_0$ and $\ell_3$ provide order-of-magnitude estimates of peeling velocity and size of disturbed region, Fig.~\ref{fig4}c,d.

However, the system exhibits a more complex behavior that depends on a delicate interplay not only of bond transport and reactions, but also of mechanics through $h$, Eqs~(\ref{GEqn1},\ref{kon},\ref{koff}). This interplay controls the dynamical organization in the moving process zone, Supp.~Fig.~3, and ultimately front velocity. As in the diffusion-dominated case, we found that dynamics are faster for larger mechanical driving force $\theta$. However, here $\hat{v}$ is nearly proportional to the vertical component of force ($\sin\theta$), Fig.~\ref{fig4}d(inset). We finally examined the influence of the slip bond effect, finding a very strong increase of $\hat{v}$ as bonds became more sensitive to force, i.e.~when $f_\beta$ becomes smaller than the force-scale close to the edge given by $f_\gamma$, Fig.~\ref{fig4}e. 

In summary, unlike the non-local and self-stabilizing peeling dynamics of the diffusion-dominated regime, here the front moves at a constant speed that depends only on the driving force and material parameters $\hat{v}(\theta; D_i, k, f_\beta, \bar{k}_{\rm off}, \bar{k}_{\rm on})$ akin to the kinetic law for the motion of a material interface \cite{Abeyaratne2006-kd}. The unconventional tear-out described here also differs from the classical tear-out for immobile bonds \cite{Dembo1988} in that it fundamentally depends not only on marginal unbinding but also on small-scale diffusion near the front.

\subsection*{Reaction-diffusion regime}

To examine an intermediate regime, we kept the off-rate  $\bar{k}_{\rm off} = 10$ s$^{-1}$ representative of weak bonds such as cadherins and considered diffusion constants typical of adhesion molecules on lipid membranes, $D_2 = 2D_1= 0.5$~$\upmu$m$^2$/s. In this regime, depending on the magnitude of the applied force, the adhesive patch can either fail or reach a stable configuration with full uniformization and equilibration of the chemical potentials of bonds and free binders, Supp.~Movie 4. With our choice of $\bar{k}_{\rm off}$, reactions take place much faster than diffusion, $\tau_{\rm reac} \approx 0.1\,{\rm s} \ll \tau_{\rm diff,1}\approx 25$ s, and thus chemical potentials between bonds and free binders locally equilibrate very quickly in the adhesion patch, Suppl.~Fig.~4a. For longer-lived bonds, e.g.~$\bar{k}_{\rm off} = 0.1$ s$^{-1}$, reactions and diffusion dynamically compete, Suppl.~Fig.~4b, leading to stronger adhesions as a whole, Suppl.~Fig.~4c,d.

Going back to the case of fast reaction rates  ($\bar{k}_{\rm off} = 10$ s$^{-1}$) and first focusing on compliant ideal bonds ($f_\beta = +\infty$), we studied in detail the peeling dynamics, which exhibit multiple scales in space and time as shown the kymographs, Fig.~3b-e. At short times commensurate to the diffusion time-scale in the patch $\tau_{\rm diff,1}$, the dynamics proceed similarly to self-similar diffusion-dominated regime, Fig.~3d,f(inset), but now, as in the reaction-dominated regime, fast unbinding localizes at the edge of the patch to reduce the chemical potential of stretched and concentrated bonds, Fig.~3e, leading to a fast initial decrease of the total number of bonds, Fig.~3g(inset). Examining early times, $t=t_1, t_2$, even if $\mu_1$ and $\mu_2$ equilibrate and uniformize within the adhesion patch, the local excess chemical potential of free binders $\mu_2$ resulting from fast unbinding has not had time to equilibrate in the rest of the vesicle, driving diffusion of free binders away from the patch, Fig.~3a,b, thereby reducing $\mu_2$ in the patch and further driving unbinding reactions, Fig.~3g. This process is much slower since it is controlled by the diffusive time-scale over the vesicle, $\tau_{\rm diff,2}= 100 \tau_{\rm diff,1}$. Since $\tau_{\rm diff,2}\gg\tau_{\rm diff,1}\gg \tau_{\rm reac}$, now reactions take place nearly uniformly in a quasi-equilibrated adhesion patch. With fewer bonds, mechanical equilibrium at the interface requires reducing the size of the patch, which decreases the contact angle and increases bond concentration, Eq.~(\ref{mech_int}). In turn, the higher bond concentration favors further unbinding. Thus, the much slower dynamics during this second phase is complex, multiphasic, and depend on the diffusion of free binders over the entire vesicle with time-scale $\tau_{\rm diff,2}$, Fig.~3f,g.

The speed and outcome of these dynamics depend on the magnitude of  $F$. To characterize adhesion strength, we simulated the dynamics for  different forces and tracked the time to complete failure $t_{\rm fail}$, Fig.~\ref{fig:mixed}h, finding that lifetime very abruptly increases as force is reduced  \cite{PhysRevLett.123.228102}. This suggests the existence of a threshold force below which the patch is long-lived and above which decohesion occurs rapidly. Consistent with this, lifetime closely follows a power-law $t_{\rm} \propto (F-F_c)^a$ with $a\approx -2.2$ and the critical force $F_c$ a fitting parameter, Fig.~3h(inset). Thus, $F_c$ can be interpreted as the strength of the adhesion patch. This mesoscopic notion of strength should depend on the microscopic strength of individual bonds given by $f_\beta$. As in the FKPP regime, we found that force sensitivity of slip bonds only plays a significant role when $f_\beta<f_\gamma$, in which case lifetime at fixed $F$ and strength  dramatically reduce, Fig.~4h.

Up to now, our model assumes a dilute limit of molecules on the membrane. However, force application leads to increasing molecular crowding, which should affect the dynamics of decohesion by changing the interplay between reactions and diffusion. To understand this, we developed a model accounting for crowding, Suppl.~Note 1. This model shows that close to a maximum concentration of molecules,  $c_{\rm max}$, the chemical potential of bonds rapidly increases, which accelerates unbinding reactions and increases the effective diffusion coefficient. 
For high crowding ($c_{\rm max}/c_0 = 5$), our simulations show that concentrations uniformize much more rapidly than in the dilute limit to reach saturation in the patch, which rapidly shrinks until failure due to unbinding reactions taking place throughout the adhesion, Suppl.~Fig.~5. Thus, crowding favors a tear-out mechanism different from the FKPP-regime, which exhibits a highly localized front of reactions. For cadherins on lipid vesicles,  $c_{\rm max}/c_0 \approx 20$ \cite{pontani2016cis}, our model still predicts significant embrittlement caused by crowding relative to the dilute limit.

\section*{Summary and discussion}

In summary, we have developed an out-of-equilibrium model self-consistently coupling diffusion, binding and unbinding reactions and mechanics to understand the dynamics of peeling  between fluid membranes bridged by mobile adhesion molecules forming transient bonds. We have used this model to map various distinct and biologically relevant scenarios of forced decohesion amenable to experimental examination. (1) For long-lived mobile bonds, adhesion patches shrink and become concentrated in a self-stabilizing process controlled by diffusion. At short times, the system evolves according to the self-similar dynamics of a classical Stefan problem with the interface moving as $(\hat{s}(t)-\hat{s}_0)\propto \sqrt{t}$. (2) For short-lived bonds with low diffusivity, such as cadherins partially immobilized by the cytoskeleton, we have identified a new unconventional tear-out regime characterized by FKPP-like traveling solutions with $(\hat{s}(t)-\hat{s}_0)\propto t$, localized reactions in the vicinity of the interface, but also by small-scale diffusion in a process zone of size $\sqrt{D_1/ \bar{k}_{\rm off}}$. The interplay between diffusion and reactions sets the order of magnitude of the front speed $\sqrt{D_1 \bar{k}_{\rm off}}$, but this speed is strongly influenced by the applied force and by the ratio between force sensitivity $f_\beta$ and the characteristic force born by bonds close to the interface $f_\gamma =\sqrt{k\cdot k_BT}$. (3) For mobile short-lived bonds such as cadherins on a lipid membrane, the system exhibits a hierarchy of reaction and diffusion time-scales resulting in multi-phasic dynamics.
The reinforcing effect of bond motion and the weakening effect of bond breaking compete in a force-dependent manner, defining the strength of the patch below which peeling arrests and above which peeling rapidly leads to complete failure. Strength strongly decreases for sensitive bonds ($f_\beta <\sqrt{k\cdot k_BT}$) and with molecular crowding.

Although our minimal model ignores important aspects of cell-cell adhesion, the physical rules identified here should bear biological relevance. We have shown how the ability of bonds to laterally move in fluid-fluid adhesive interfaces leads to very rich repertoire of peeling scenarios that cells can use to stabilize cell-cell junctions during physiological stretch, or to selectively detach during morphogenesis. For instance, cells can effectively tune adhesive strength, and hence their ability to stay adhered or disengage, by controlling molecular properties of bonds such as stiffness $k$ and force sensitivity $f_\beta$, e.g.~through extracellular Ca$^{2+}$, by controlling the number of transmembrane crowding molecules, or by controlling the actively generated surface tension. Beyond cells, our study also provides a conceptual framework for artificial biomimetic systems with a comparable degree of adhesive tunability \cite{Parolini:2015aa}.

\subsection*{Acknowledgements}
The authors acknowledge the support of the European Research Council (CoG-681434), the European Commission (Project No. H2020-FETPROACT-01-2016-731957), the Spanish Ministry for Science and Innovation (PID2019-110949GB-I00), and the Generalitat de Catalunya (2017-SGR-1278 and ICREA Academia prize for excellence in research). IBEC and CIMNE are recipients of a Severo Ochoa Award of Excellence from the MINECO.

\bibliographystyle{my_unsrtnat_bis}
\bibliography{Bibliography}

\end{document}